\PassOptionsToPackage{unicode}{hyperref}
\PassOptionsToPackage{hyphens}{url}
\documentclass[
]{article}
\usepackage[margin=1.25in]{geometry}
\usepackage{amsmath,amssymb}
\usepackage{iftex}
\ifPDFTeX
  \usepackage[T1]{fontenc}
  \usepackage[utf8]{inputenc}
  \usepackage{textcomp} 
\else 
  \usepackage{unicode-math} 
  \defaultfontfeatures{Scale=MatchLowercase}
  \defaultfontfeatures[\rmfamily]{Ligatures=TeX,Scale=1}
\fi
\usepackage{lmodern}
\ifPDFTeX\else
\fi
\IfFileExists{upquote.sty}{\usepackage{upquote}}{}
\IfFileExists{microtype.sty}{
  \usepackage[]{microtype}
  \UseMicrotypeSet[protrusion]{basicmath} 
}{}
\makeatletter
\@ifundefined{KOMAClassName}{
  \IfFileExists{parskip.sty}{%
    \usepackage{parskip}
  }{
    \setlength{\parindent}{0pt}
    \setlength{\parskip}{6pt plus 2pt minus 1pt}}
}{
  \KOMAoptions{parskip=half}}
\makeatother
\usepackage{xcolor}
\usepackage{graphicx}
\makeatletter
\def\maxwidth{\ifdim\Gin@nat@width>\linewidth\linewidth\else\Gin@nat@width\fi}
\def\maxheight{\ifdim\Gin@nat@height>\textheight\textheight\else\Gin@nat@height\fi}
\makeatother
\setkeys{Gin}{width=\maxwidth,height=\maxheight,keepaspectratio}
\makeatletter
\def\fps@figure{htbp}
\makeatother
\ifLuaTeX
  \usepackage{selnolig}  
\fi
\IfFileExists{bookmark.sty}{\usepackage{bookmark}}{\usepackage{hyperref}}
\IfFileExists{xurl.sty}{\usepackage{xurl}}{} 
\hypersetup{
  hidelinks,
  pdfcreator={LaTeX via pandoc}}

\author{}
\date{}

\begin{document}

{\Large \bf Rotational 3D printing of active-passive filaments and lattices with
programmable shape morphing}

\vspace{2em}

Mustafa K. Abdelrahman\textsuperscript{1}, Jackson K.
Wilt\textsuperscript{1}, Yeonsu Jung\textsuperscript{1}, Rodrigo
Telles\textsuperscript{1}, Gurminder K. Paink\textsuperscript{1},
Natalie M. Larson\textsuperscript{1,2}, Joanna
Aizenberg\textsuperscript{1,3}, L. Mahadevan\textsuperscript{1,4,5*},
and Jennifer A. Lewis\textsuperscript{1,6*}

\textsuperscript{1}Harvard John A. Paulson School of Engineering and
Applied Sciences, Harvard University, Cambridge, MA, USA

\textsuperscript{2}Department of Mechanical Engineering, Stanford
University, Stanford, CA, USA

\textsuperscript{3}Department of Chemistry and Chemical Biology, Harvard
University, Cambridge, MA, USA

\textsuperscript{4}Department of Physics, Harvard University, Cambridge,
MA, USA

\textsuperscript{5}Department of Organismic and Evolutionary Biology,
Harvard University, Cambridge, MA, USA

\textsuperscript{6}Wyss Institute for Biologically Inspired Engineering,
Harvard University, Cambridge, MA, USA

*Corresponding Authors: L. Mahadevan and Jennifer A. Lewis

\textbf{Email:}
\href{mailto:lmahadev@g.harvard.edu}{\nolinkurl{lmahadev@g.harvard.edu}};
\href{mailto:jalewis@seas.harvard.edu}{\nolinkurl{jalewis@seas.harvard.edu}}

\textbf{Author Contributions:} M.K.A., J.K.W., N.M.L., Y.J., L.M., and
J.A.L. designed the research. M.K.A., J.K.W., R.T., and G.K.P. performed
the experimental research. M.K.A., J.K.W., R.T., and G.K.P. analyzed the
data. Y.J. and L.M. constructed the theoretical framework. Y.J.
performed the computer simulations. M.K.A., J.K.W., Y.J., R.T., G.K.P.,
N.M.L., J.A., L.M., and J.A.L. wrote the manuscript. All authors
reviewed and approved the final manuscript.

\textbf{Competing Interest Statement:} The authors declare no competing
financial interests.

\textbf{Classification:} Physical Sciences; Applied Physical Sciences


\textbf{This PDF file includes:}

\begin{quote}
Main Text

Figures 1 to 5
\end{quote}

\clearpage

\textbf{Abstract}

Natural filaments, such as proteins, plant tendrils, octopus tentacles,
and elephant trunks, can transform into arbitrary three-dimensional
shapes that carry out vital functions. Their shape-morphing behavior
arises from intricate patterning of active and passive regions, which
are difficult to replicate in synthetic matter. Here, we introduce a
filament-centric strategy for programmable shape morphing in which
intrinsic curvature and twist are directly encoded within multimaterial
elastomeric filaments during fabrication. By harnessing rotational
multimaterial 3D printing (RM-3DP), we directly prescribe the filament's
natural curvature--twist field $\mathbf{k}(s)$ through controlled material
distribution and helical liquid crystal mesogen alignment. When heated
above their nematic-to-isotropic transition temperature
($T_\text{NI}$), the helically aligned LCE regions contract along
their local director field, while passive regions remain essentially
unchanged. This approach enables independent control of bending and
torsion at every cross-section along the filament centerline: the
principal natural curvatures of the filament along two orthogonal axes
as well as the local twist. Next, we printed architected lattices
composed of unit cells formed by sinusoidal filaments that either
reversibly contract, expand, or exhibit out-of-plane deformations.
Discrete elastic rod simulations of Janus filaments with different
natural curvatures and twist, which are interconnected within the
printed lattices, allow accurate prediction of their observed
shape-morphing behavior. By integrating active-passive elastomers,
additive manufacturing, and computational modeling, we have created
shape-morphing matter with complex programmable responses for
applications that rely on adaptive, robotic, or deployable
architectures.

\textbf{Significance Statement}

Natural filaments have exceptional control over curvature and twist
enabled by directional responses embedded within their internal
structure. To emulate this complex behavior, we use rotational
multimaterial 3D printing (RM-3DP) to create composite fibers composed
of active liquid crystal elastomers (LCEs) and passive elastomers. By
controlling the rotation rate on-the-fly during printing, one can
fabricate Janus filaments and lattices with spatially programmable
composition, alignment, and shape-morphing behavior, including
reversible bending, coiling, and twisting when thermally cycled above
and below their nematic-to-isotropic transition temperature. A
theoretical and computational framework corroborates our experimental
findings and paves the way for a quantitative framework to understand
and design shape-morphing filaments and lattices.

\clearpage

\textbf{Main Text}

\textbf{Introduction}

Nature is replete with active filaments, e.g., proteins (1,2), plant
tendrils (3--5), octopus tentacles (6), and elephant trunks (7--9) that
exhibit remarkable transformations from straight forms to curved and
twisted geometries that fulfill vital biological functions. Proteins,
for instance, fold from a random coil into a specific three-dimensional
structure to achieve biological activity. Similarly, plant tendrils
helically coil to provide structural support, allowing plants to climb
toward sunlight, while octopus tentacles and elephant trunks bend,
twist, and curl to manipulate objects and facilitate communication.
These systems illustrate the intimate coupling between filament geometry
and function, arising from the deterministic arrangement of their
constituents. For instance, the precise sequence of amino acids
determines protein folding pathways, stiff lignified cells drive tendril
coiling (10,11), and the patterned activation of muscles combined with
passive strain asymmetry enables the versatile movement of the octopus
tentacle or the elephant trunk (9,12). Inspired by these natural
examples, synthetic filaments have been developed that exhibit
shape-morphing responses for applications in adaptive materials
(13--15), soft robotics (16--18), and biomedical devices (19--21).

The shape-morphing behavior of soft filaments can be programmed either
extrinsically via differential deformation patterning, intrinsically
through the incorporation of stimuli-responsive polymers, or a
combination thereof. Extrinsically programmed filaments, such as
bilayers with varying coefficients of thermal expansion, yield
structures that primarily exhibit bending deformations (22--24). More
complex deformations, such as twisting and coiling, can be achieved by
spatially patterning filaments with inhomogeneous deformations, such as
hydrogels with differential swelling ratios (25--28) or elastomers with
varying coefficients of thermal expansion (29). Furthermore, by coupling
multiple filaments with disparate responses, one can achieve
out-of-plane deformations. Indeed, from a mathematical perspective, at
every cross-section, there are three translational and three
orientational degrees of freedom for the material axes attached to the
centerline. The translational degrees correspond to the axial stretch
and two transverse shears, while the orientational degrees are
associated with two curvatures and a twist. Together, these degrees of
freedom span the Euclidean group SE(3) (30,31). For slender inextensible
filaments, the translational degrees of freedom are difficult to actuate
(owing to their large stiffnesses), but the orientational degrees of
freedom are relatively easy to actuate, resulting in predominantly
rotational deformations, i.e., SO(3) (31). From a practical perspective,
pneumatic actuators composed of inflatable hollow tubes can exhibit
complex 3D deformations even at the filamentary level. Soft robotic
matter can be rapidly created by soft lithography (32,33) or bubble
casting (34), which enable grasping (35), trajectory matching (36), and
locomotion (33). However, the reliance on tethered pneumatic systems
limits their applicability.

Untethered soft materials that exhibit shape deformation can be achieved
using intrinsically programmed, stimuli-responsive materials. For
example, the shape memory effect of polymer networks has enabled the
development of self-tying filaments; however, shape deformation is
typically not reversible, and shape selection is constrained by the need
for a mechanical programming step post-synthesis (21). In contrast,
liquid crystal elastomers (LCEs) enable reversible, untethered actuation
without requiring a mechanical programming step (37). LCEs are
stimuli-responsive polymers that undergo reversible shape
transformations for at least 10$^6$ cycles (38). At the
molecular level, LCEs contain rod-like molecules, known as mesogens,
that undergo a transition from a nematic (ordered) state to an isotropic
(disordered) state upon heating above their nematic-to-isotropic phase
transition temperature ($T_\text{NI}$), resulting in large,
reversible changes in shape (39). The orientation of the nematic
director can be programmed through surface alignment (40), magnetic
alignment (41--43), or shear-induced alignment (44). While surface and
magnetic alignment enable precise programming of structures that bend
and twist, they are limited to thin films (45) (\textasciitilde50 µm) or
microscopic structures (100 -- 500 µm), respectively (41). Direct ink
writing (DIW) offers a facile method for printing LCE filaments with
programmable, shear-induced alignment for applications ranging from
textiles (46--49) to self-sensing artificial muscles (50--52). However,
monolithic LCE filaments are typically limited to contractile actuations
along the printing (i.e., alignment) direction. To encode more complex
deformations, such as coiling, one must introduce additional processing
steps during their fabrication (53--55). More broadly, a direct method
to prescribe the intrinsic curvature and twist of individual filaments
remains largely unexplored.

Here, we harness rotational multimaterial 3D printing (RM-3DP) (56) to
fabricate architected elastomer filaments and lattices with programmable
shape-morphing behavior. Using customized nozzles with two semi-circular
channels, active LCE and passive acrylate elastomer inks are co-extruded
to produce Janus filaments with sharply defined internal interfaces
(Fig. 1A). Nozzle rotation during printing imposes a helical LCE mesogen
orientation and systematically varies the spatial distribution of the
two materials along the filament. When heated above $T_\text{NI}$,
the active LCE regions contract along the nematic director, while the
passive elastomer regions remain largely unchanged, allowing encoding of
myriad shape transformations within architected filaments and lattices.
Unlike traditional approaches that rely on dense filament networks or
thin sheets that morph through differential growth and encode bending
through the first and second fundamental forms (25,28,29), RM-3DP
operates in the filamentary limit and treats each filament as an
independent Cosserat rod, allowing direct control over local curvature
and twist via prescribed material orientation. Similar to dense and
sheet-based systems, RM-3DP can generate smooth, continuous curvature,
but it also enables precise control over torsion, multiaxial coupling
(i.e., where bending and twisting occur simultaneously), and
programmable anisotropy. As simple demonstrations of their potential
applications, we fabricated active filters and grippers capable of
simultaneously manipulating multiple objects.

\textbf{Results and Discussion}

Janus filaments are fabricated by co-extruding an oligomeric LCE ink, in
which rigid mesogens are incorporated along the polymer backbone, and a
soft acrylate ink via RM-3DP. A custom nozzle, containing two
semi-circular channels with a total nozzle diameter of 1 mm, is used
during printing (Supplementary Fig. 1). The liquid crystal mesogens in
the active elastomer ink exhibit shear-induced director alignment when
printed in the nematic phase (Fig. 1A). During printing, the nozzle can
be rotated to precisely position each material along the filament length
(Fig. 1B; Supplementary Movie 1). The active and passive elastomer inks
exhibit shear thinning behavior at 25 $^\circ$C, the printing temperature (Fig.
1C; Supplementary Fig. 2). Both inks contain acrylate-terminated
oligomers ensuring their strong adhesion after UV curing, as confirmed
by a 90° peel test. Their peak adhesion strength is 1289 $\pm$ 334 N
m$^{-1}$ (Supplementary Fig. 3). Cohesive failure of the
weaker (passive) elastomer indicates that the strength of adhesion
between the two elastomers exceeds that of the passive material. The
active and passive elastomers exhibit disparate bulk actuation and
mechanical properties. When heating the pure controls above
$T_\text{NI}$, the active LCE filaments undergo a pronounced
contraction (57), while the passive filaments remain largely unchanged
(Fig. 1D; Supplementary Fig. 4). Normalized lengths of 0.67 $\pm$ 0.02 and
1.03 $\pm$ 0.02 are observed for the purely active and passive filaments,
respectively, upon heating to 150 $^\circ$C (Fig. 1E). The elastic modulus of
purely active LCE filaments in the polydomain state is 29.67 $\pm$ 2.55 MPa
(T\textsubscript{g} = -2.71 $\pm$ 0.73 $^\circ$C), while that of the purely passive
filaments is 0.57 $\pm$ 0.07 MPa (T\textsubscript{g} = -63.12 $\pm$ 0.47 $^\circ$C) at
25 $^\circ$C (Supplementary Figs. 5--6). The observed 50-fold difference in
elastic moduli between active and passive elastomer provides an
additional design parameter for programming filament curvature.
Importantly, printed filaments exhibit larger bending responses when the
elastic modulus ratio between these features is increased (Supplementary
Fig. 7).

We encode shape-morphing behavior by printing active-passive elastomeric
filaments with controlled twist and pitch (Fig. 2A). In this framework,
the primary design variable is not merely material placement, but the
spatially varying natural curvature--twist vector $\mathbf{k}(s)$, which defines
the intrinsic geometry of the filament independent of external loads.
The local material frame
\(Q(s) = \{ d_{1}(s),d_{2}(s),d_{3}(s)\}\ \)defines the filament
orientation, with \(d_{3}(s)\ \)tangent to the centerline and
\(\left\{ d_{1},d_{2} \right\}\ \)spanning the cross-sectional plane.
The interfacial normal vector \(n(s)\) lies within this plane and
rotates along the filament according to the prescribed angle
\(\alpha(s)\). To model this, we represent each filament as a continuous
3D curve, \(r(s)\), and introduce a discrete elastic rod model (Fig. 2B)
(58,59). Each discrete filament section consists of a Janus
cross-section defined by a normal vector, \(n(s)\), orthogonal to the
interface between the active and passive domains in the cross-sectional
plane spanned by \(\{ d_{1},d_{2}\}\). The orientation of this interface
is described by an angle \(\alpha(s)\), such that
\(n(s) = \cos(\alpha)d_{1}(s) + \sin(\alpha)d_{2}(s)\), and its spatial
rotation \(d\alpha/ds\) is directly prescribed through the nozzle
rotation rate, \(\omega\), through the kinematic relation
\(\omega = vd\alpha/ds\), where \(v\) is the translation speed (Fig.
2C,D; Supplementary Movie 1). A filament with constant \(\alpha\) yields
a Janus filament, whereas a filament with varying
\(\alpha(s)\ \)introduces a programmed interfacial twist \(d\alpha/ds\)
along the filament length.

Each filament is modeled as a slender, inextensible, and unshearable rod
of length \(L\) with a circular cross-section of diameter \(a\). Their
deformation can then be described in terms of the relative orientation
of each cross-section as a function of location along its axis, and its
configuration may be described by the centerline, \(r(s)\), and the
material frame \(\{ d_{1}(s),d_{2}(s),d_{3}(s)\}\). The elastic energy
is then expressed as:

\[\epsilon = \frac{1}{2}\int_{0}^{L}{\left( k - \overline{k} \right)^{T}B(k - \overline{k}})ds = \frac{1}{2}\int_{0}^{L}\left\lbrack B\left( \left( \kappa_{1} - {\bar{\kappa}}_{1} \right)^{2} + \left( \kappa_{2} - {\bar{\kappa}}_{2} \right)^{2} \right) + C\left( \tau - \bar{\tau} \right)^{2} \right\rbrack ds\]

where \(\mathbf{k}(s) = \{\kappa_{1}(s),\kappa_{2}(s),\tau(s)\}\) is the
curvature-twist vector,
\(\bar{\mathbf{k}}(s) = \{{\bar{k}}_{1}(s),{\overline{k}}_{2}(s),\bar{\tau}(s)\}\)
is the natural curvature induced by differential thermal strains or
growth, and \(B = diag(B,B,C)\). Here, the bending stiffness
\(B = EI = E(\frac{\pi a^{4}}{4})\) and the twisting stiffness
\(C = GJ = G(\frac{\pi a^{4}}{2})\), where \(E\) and \(G\) are the
Young's modulus and shear modulus, respectively. The equilibrium shape
minimizes $\epsilon$; under free boundary conditions, the curvature and twist
relax to their natural values, \(\mathbf{k}(s) = \bar{\mathbf{k}}(s)\) and
\(\tau(s) = \bar{\tau}(s)\), such that bending occurs along the
programmed interfacial normal \(n(s)\). We note that our 3D rotational
printing platform allows one to control the vector
\(\bar{\mathbf{k}}(s) = \{{\bar{k}}_{1}(s),{\overline{k}}_{2}(s),\bar{\tau}(s)\}\)
by varying the rate of rotation and the relative ratio of
active-to-passive elastomers, providing complete control over the
spatial patterning of the orientational degrees of freedom of the
filamentary structures.

The dimensionless rotation rate, \(\omega^{*}\), quantifies the degree
of interfacial twist imposed during RM-3DP, via
\(\omega^{*} = \frac{Rd\alpha}{ds} = \frac{Rd\alpha}{dt}\frac{dt}{ds} = \frac{R\omega}{v}\).
Filaments printed without rotation (\(\omega^{*}\) = 0) form straight
Janus filaments that purely bend upon heating due to differential
thermal contraction between the active and passive halves, inducing
filament curvature (60). The magnitude of curvature scales with the
Weissenberg number, \(Wi = \lambda\dot{\gamma}\), which reflects
shear-induced mesogen alignment during extrusion (61). For a fixed
nozzle geometry and filament cross-section, increasing the print speed
\(v\) increases the volumetric flow rate (\(Q \propto v\)), thereby
increasing the shear rate and hence \(Wi\). Accordingly, as \(v\)
increases from 0.25 to 3.0 mm s$^{-1}$, the order parameter
\(S\) rises from 0.150 $\pm$ 0.002 to 0.230 $\pm$ 0.003, leading to stronger
contraction and greater curvature (Supplementary Figs. 8--9) (62).
Consequently, the maximum curvature of the actuated filament can be
tuned from 0.13 $\pm$ 0.02 mm$^{-1}$ (for \(v\) = 0.25 mm
s$^{-1}$) to 0.23 $\pm$ 0.01 mm$^{-1}$ (for \(v\)
= 0.5 mm s$^{-1}$) to 0.47 $\pm$ 0.05 mm$^{-1}$
(for \(v\) = 1.75 mm s$^{-1}$) to 0.59 $\pm$ 0.04
mm$^{-1}$ (for \(v\) = 3 mm s$^{-1}$) (Fig.
2D; Supplementary Fig. 10). Notably, under repeated thermal cycling
between 25 $^\circ$C and 175 $^\circ$C, the filaments exhibit highly reversible
curvature with no observable degradation over 100 cycles (Supplementary
Fig. 11). No interfacial debonding, delamination, or slip is observed,
even at high curvature, which we attribute to covalent bonds that form
between the two acrylate-based (active and passive) elastomer inks.

Programming a spatially varying interfacial twist, \(\alpha(s)\),
enables control over the twist and out-of-plane bending (or torsion) of
the printed filament upon heating above $T_\text{NI}$
(Supplementary Movie 1). Quantitatively, the dimensionless rotation
rate, \(\omega^{*}\), controls both the pitch and molecular orientation
of the LCE filament. As confirmed by two-dimensional (2D) WAXS
measurements across a filament cross-section, rotational printing
imposes a well-defined helical mesogen orientation, whose angle is given
by $\varphi = \tan^{-1}$($\omega^*$) (63) (Fig. 2E; Supplementary
Fig. 12). Pure LCE (active) filaments inherit this helical director
field uniformly across their cross-section and display a characteristic,
continuous progression in deformation modes with increasing
$\omega^*$. At low $\omega^*$, shallow
helical angles produce bending-dominated morphologies with the largest
decrease in end-to-end length ($L/L_0$ = 0.80 $\pm$ 0.03 at
$\omega^*$ = 0.06) (Fig. 2F). As
$\omega^*$ increases, the nematic director rotates,
introducing a circumferential contraction component, yielding mixed
bending--twisting shapes and a corresponding increase in end-to-end
length ($L/L_0$ = 0.86 $\pm$ 0.01 at $\omega^*$ = 0.25). Twist
reaches its maximum magnitude at $\omega^*$ = 1, where
the director approaches a helical angle of 45° relative to the
filament's centerline and optimally converts uniaxial contraction into
torsional strain (64), producing nearly twist-dominated helices with
minimal end-to-end length shortening ($L/L_0$ = 0.91 $\pm$ 0.01 at
$\omega^*$ = 1.0). Further increasing
$\omega^*$ increases the helical angle towards 90°,
causing bending to reemerge, leading to end-to-end length shortening
($L/L_0$ = 0.87 $\pm$ 0.02 at $\omega^*$ = 2). This
``parabolic'' variation in normalized length highlights the intrinsic
deformation progression of pure LCE filaments upon heating; however,
introducing a passive region within the cross-section fundamentally
alters this shape-morphing behavior.

Composite filaments inherit the same helical director field but only
half of the cross-section is active, resulting in altered shape-morphing
behavior. At low $\omega^*$, the active region lies
predominantly on one side of the filament, establishing a stable
cross-sectional strain gradient that strongly favors bending and
produces coiled filaments with large end-to-end length shortening ($L/L_0$
= 0.22 $\pm$ 0.01 at $\omega^*$ = 0.06) (Fig. 2G;
Supplementary Fig. 10, Supplementary Movie 2). As
$\omega^*$ increases, the helical angle increases,
producing morphologies with coupled bending and twisting, including
toroidal coils ($\omega^*$ = 0.25), reminiscent of DNA
supercoiling. However, at high $\omega^*$, rapid
variation between the active and passive elastomers emerges along the
filament's length, eliminating the geometric condition required for
bending. As a result, a stable bending axis cannot form and bending is
suppressed. Therefore, twisting becomes the dominant mode of
deformation, producing tightly wound helices with nearly no end-to-end
length shortening ($L/L_0$ = 0.98 $\pm$ 0.02 at $\omega^*$ =
2). This transition from bending-dominated to twisting-dominated
morphologies with increasing $\omega^*$ can be
rationalized through the Kirchhoff analogy, which links elastic filament
statics to spinning-top dynamics (30,31), where rapid rotation
stabilizes the axis of motion through torsional coupling (65).

The ability to program twisting and coupled bending-twisting
deformations could find potential applications as filamentary grippers
(16) or synthetic assemblies (13). In each case, the ability to
precisely program filament shape is critical for achieving complex,
emergent functionality. Furthermore, beyond purely mechanical actuation,
emergent functionalities may also be achieved by replacing the passive
elastomer with functional materials. For example, incorporating
pressure-sensitive adhesive elastomers could enable adaptive gripping,
while conductive polymers could impart sensing capabilities.
Miniaturization of these filaments could enable further applications,
such as colloidal robots (66) or artificial cilia (67). In this work,
nozzle sizes are limited due to the resolution limits of our DLP resin
printer (\textasciitilde{} 50 µm). Nonetheless, by reducing the nozzle
size from 1 mm to 0.5 mm, filament diameters decrease from 600 µm to 300
µm (Supplementary Fig. 13). However, one caveat is that they must be
printed at lower speeds (0.5 mm s$^{-1}$), which reduces
the desired shear-induced LCE alignment. We note that although it would
be feasible to create nozzles that are 0.1 mm in outer diameter using a
higher resolution 3D printer, one must concomitantly reduce the ink
viscosity to ensure high printing speed to encode the desired LCE
alignment. With this understanding of how printing parameters govern 3D
deformation, we next sought to couple rotation with 2D print paths to
create spatially patterned, morphable architectures.

Because intrinsic curvature is encoded at the filament level, geometric
programming propagates hierarchically to the lattice scale, where
collective filament interactions give rise to emergent shape
transformations. At the filament level, the active LCE elements contract
along the direction of director alignment and expand in the transverse
direction upon heating above their nematic-to-isotropic transition
temperature (Supplementary Fig. 14). By coupling the aligned LCE
elements to passive elements within these Janus filaments, their
differential actuation response gives rise to bending. In particular,
when the composite filament is printed with an initial curvature and the
LCE layer is positioned on the outer (longer) side of the curvature,
axial contraction of the LCE reduces the arc-length mismatch between the
active and passive layers, leading to filament straightening. As a
benchmark, we printed lattices with sinusoidal filaments arranged into
homogeneous unit cells, as this geometry enables positioning of the
active elastomer on either the outer or inner side of the curvature to
drive expansion or contraction (Fig. 3A,B). When the active elastomer is
placed on the outer curvature, asymmetric contraction upon heating
decreases the filament curvature, causing it to straighten and increase
its end-to-end distance (Fig. 3C). Conversely, positioning the active
elastomer on the inner curvature induces an increase in curvature upon
heating, shortening the filament's end-to-end distance (Fig. 3D). Upon
heating, the normalized end-to-end lengths reach 1.37 $\pm$ 0.07 for
expanding filaments and 0.64 $\pm$ 0.05 for contracting filaments
(Supplementary Fig. 15). Both deformation modes are fully reversible
upon cooling.

Extending this concept, we fabricated 4×4 lattices composed entirely of
either expanding or contracting filaments, yielding lattices that expand
or contract uniformly upon heating (Fig. 3E,F; Supplementary Movie 3).
Expanding lattices exhibit an increase in area of 99 $\pm$ 11\%, whereas
contracting lattices decrease in area by 28 $\pm$ 11\% (Fig. 3G;
Supplementary Movies 4,5). For contracting lattices, deformation is
limited by self-contact between neighboring filaments, which leads to a
mechanically jammed state that inhibits further contraction and results
in a reduced actuation magnitude. We note that adhesion interactions
were not observed in these lattices, enabling reversibility over
repeated heating and cooling cycles. After ten heating and cooling
cycles between 25 $^\circ$C and 150 $^\circ$C, contracting lattices return to a
normalized area of 1.07 $\pm$ 0.03 of their original state, while expanding
lattices recover to 1.19 $\pm$ 0.05 (Fig. 3H; Supplementary Figs. 16--17).
Having demonstrated expansion and contraction as independent modes, we
next explored spatially combining these responses within a single
lattice.

By coupling expanding and contracting regions within the same lattice
design, one can drive out-of-plane deformations akin to biological
morphogenesis. In nature, the patterning of differential growth or
contraction enables the formation of complex 3D structures, such as the
folding of organs (68)~ and the curling of plant tissues (69).
Similarly, by architecting lattices composed of unit cells with
non-trivial \(\alpha(s)\), 3D shape transformations arise from the
temperature-driven changes in the natural curvature of sinusoidal
filaments. These curvature variations generate configurations that
minimize the elastic bending energy \(\epsilon\) upon heating, producing
out-of-plane morphologies with tunable positive or negative gaussian
curvature. Inspired by these principles, we numerically modeled and
experimentally validated heterogeneous lattices composed of both
expanding and contracting filaments (Supplementary Fig. 18). We modeled
these lattices as networks of discrete elastic rods, where each
sinusoidal filament connects four branch nodes to form a repeating unit
cell. At a branch node where multiple sinusoidal edges meet, the total
bending and twisting energies are computed by summing the contributions
from all unique pairs of connecting edge segments (70).

In our architected lattice composed of sinusoidal filaments, the
effective elongation and contraction are programmed by the direction of
bending: elongation corresponds to a reduction of bending curvature
along the end-to-end direction, while contraction corresponds to an
increase. This mismatch prevents purely in-plane deformation. Instead,
the structure undergoes out-of-plane deformation, minimizing the energy
associated with bending in the original end-to-end direction at the
expense of increased transverse bending and twist (71). As a simple
demonstration, two representative configurations are designed to realize
opposite curvature behaviors. In the first, expansive filaments are
positioned at the lattice center and contractive filaments along the
perimeter. Both simulations and experiments show that this arrangement
produced spherical morphologies with positive gaussian curvature, as
initially flat lattices rose into dome-like shapes upon actuation (Fig.
4A--C; Supplementary Movie 6). In contrast, inverting the pattern by
placing contractive filaments at the center and expansive filaments at
the perimeter yielded saddle-shaped morphologies with negative gaussian
curvature, again confirmed through simulations and experiments (Fig.
4D--F; Supplementary Movie 7). We note that the evolution of the
characteristic curvature, defined as \(\sqrt{|k_{1}k_{2}|}\), where
\(k_{1}\ \)and \(k_{2}\) are the principal surface curvatures, closely
follows the measured changes in local edge lengths (Supplementary Fig.
18). Minor deviations arise from gravitational effects in the initially
flat lattices, which require a finite temperature increase to overcome
gravity and induce nonzero curvature. These gravitational effects are
more pronounced in the negatively curved lattice due to its lower
effective stiffness under vertical load, in contrast to the greater
structural rigidity of domes with positive gaussian curvature.
Nonetheless, by spatially integrating expanding and contracting
filaments within a single architecture, rather than restricting
deformation to one mode to generate curvature (72), our approach enables
reversible out-of-plane shape transformations of lattices with highly
programmable morphologies.

Unlike prior demonstrations of lattices with programmable shape
morphing, where bending arises in struts composed of multiple materials
with mismatched properties (29,73,74), this current embodiment
integrates active and passive materials within a single filament. Here,
deformation follows from a prescribed natural curvature and twist field
rather than from strain mismatch fixed by layered strut geometry,
eliminating the need to pattern bilayer or trilayer architectures.
Through rotational co-printing of active liquid crystal elastomer and
passive elastomer within one cross-section, we directly encode local
natural curvature and twist along the filament centerline. This enables
large, reversible in- and out-of-plane deformations by coupling the
intrinsic actuation strain of LCEs with precisely positioned passive
domains. Because the active and passive regions are integrated during
printing, multimaterial lattices with programmable shape morphing are
also realized in a single fabrication step, and the mechanical
description correspondingly shifts from growing elastic surfaces (27,28)
to actively deforming elastic filaments. Ultimately, rotational
multimaterial printing redefines where deformation is programmed by
shifting it from patterned multilayer strut assemblies to a single,
continuously programmed filament, thereby enabling precise control over
lattice-scale deformation and its functional consequences.

This precise control of unit cell deformation enables soft robotic
matter capable of manipulating objects in myriad ways. Active-passive
lattices composed of unit cells programmed to expand upon heating can
act as filters (Fig. 5A,B; Supplementary Movie 8). At low temperatures
below $T_\text{NI}$, the filter is in the closed state, where the
unit cell aperture is smaller than the object diameter, enabling the
lattice to catch the object. Upon heating above $T_\text{NI}$, the
lattice transitions to the open state, increasing the aperture size and
releasing the object. This reversible closed-open transition enables
on-demand capture and release. Importantly, active-passive lattices can
filter objects of varying size and geometry, such as spheres (Fig.
5C,D). Beyond filtering, architected lattices can also transfer multiple
objects simultaneously. For example, active-passive lattices programmed
to contract upon heating can function as pick-and-place tools capable of
transferring multiple objects between predefined locations (Fig. 5E-G;
Supplementary Movie 9). As a simple demonstration, acrylic rods (3.5 mm
in diameter and 6 mm in length) are transferred using a lattice-based
gripper programmed to contract. Upon heating the lattice above
$T_\text{NI}$, the unit cells contract, leading to an aperture
size smaller than the diameter of each rod, thereby gripping and
securing the objects. Once secured, the lattice along with the objects
is transferred to a second location with designated slots for placement.
Cooling below $T_\text{NI}$ causes the unit cells to expand,
increasing the aperture and releasing the objects into their prescribed
positions, where they remain even after the gripper is removed. Unlike
most soft grippers reported to date, which typically manipulate a single
object at a time, these lattices enable simultaneous pick-and-place of
multiple objects.

\textbf{Conclusion}

In summary, we establish a filament-centric strategy for programmable
shape morphing by directly encoding intrinsic curvature and twist within
multimaterial elastomeric filaments. Through rotational co-printing, we
prescribe the natural curvature--twist field $\mathbf{k}(s)$ of individual
filaments, thereby achieving independent control over the full set of
orientational degrees of freedom (i.e., the two spatially varying
curvatures and twist). This strain gradient programming drives complex
actuation via differential expansion of the constituent materials.
Furthermore, by controlling rotation and print speed along prescribed
paths, we architect lattices that exhibit heterogeneous shape morphing,
including out-of-plane actuation, consistent with computational
predictions based on discrete elastic rod theory. As demonstrations of
function, these lattices operate as dynamic filters and multi-object
grippers, illustrating how geometric control at the filament level
translates into lattice-scale mechanical performance.

More broadly, encoding curvature and twist at the filament level
represents the central advance of this work. Because deformation is
programmed geometrically rather than tied to a specific material system,
the same design principles extend beyond LCEs to other active materials,
including hydrogels, shape-memory polymers, and dielectric elastomers,
with material properties governing actuation amplitude, response time,
and durability. Our numerical framework is likewise transferable across
material classes through appropriate constitutive descriptions.
Together, our generalizable modeling and digital fabrication framework
enables the rapid design and printing of architected soft matter for
adaptive materials, soft robotics, and deployable structures.

\textbf{Materials and Methods}

\textbf{Materials:} Liquid crystal monomer,
1,4-bis-{[}4-(6-acryloyloxyhexyloxy) benzoyloxy{]}-2-methylbenzene
(RM82), was purchased from Synthon chemicals. The photoinitiator,
2,2-dimethoxy-2-phenylacetophenone (Irgacure I-651), the chain extender,
n-hexylamine, the inhibitor, butylated hydroxytoluene (BHT), and a
two-part polydimethylsiloxane (PDMS) (Sylgard 184) elastomer were
purchased from Fisher Scientific. Fumed silica, CAB-O-SIL EH-5 and
CAB-O-SIL TS-720, were purchased from Cabot. Silicone oil was purchased
from Oakwood Products (Lot No: 102516V20D). Ebecryl 8413 resin was
purchased from Allnex. Pentaerythritol tetraacrylate was purchased from
TCI Chemicals. The aliphatic urethane acrylate oligomer (CN9018) was
purchased from Sartomer. Isodecyl acrylate was purchased from
Sigma-Aldrich. Red and blue Silc Pig pigment dyes were purchased from
Smooth-On.

\textbf{Ink synthesis:} The active liquid crystal elastomer (LCE) ink
was prepared via an aza--Michael addition reaction. RM82 and
n-hexylamine were combined at a 1.4:1 molar ratio with 2 wt.\% Irgacure
I-651, 0.2 wt.\% butylated hydroxytoluene (BHT), and 5 wt.\% fumed
silica. RM82, I-651, and BHT were weighed into a 20 mL vial and heated
at 100 $^\circ$C for 1 h to melt. n-Hexylamine was then added, and the mixture
was homogenized using a heat gun and vortex mixing. Oligomerization was
carried out at 75 $^\circ$C for 18 h. After oligomerization, fumed silica (5
wt.\%) was incorporated using a planetary mixer (FlackTek, Inc.) at 2000
rpm for 10 min with 2 min rest intervals between mixing cycles. The ink
was transferred to a 10 cc UV/light-block amber syringe (Nordson EFD)
and degassed by centrifugation at 3000 rpm for 3 min (Heraeus Megafuge
8) to remove trapped air.

The passive elastomer ink was prepared from a mixture of 41.1 wt.\%
CN9018, 41.1 wt.\% isodecyl acrylate, 0.8 wt.\% Irgacure I-651, and 17
wt.\% fumed silica (CAB-O-SIL TS-720). All components except the silica
were first combined in a disposable polypropylene mixing cup. Fumed
silica was added in two portions: half was incorporated and mixed at
2000 rpm for 10 min (2 min mixing intervals with rest periods) using a
planetary mixer (FlackTek, Inc.), followed by addition of the remaining
silica and a second mixing cycle. The mixture was manually reintegrated
and subjected to a third mixing cycle to ensure homogeneity. Blue
pigment dye (0.5 wt.\%) was added prior to the final mixing step. The
ink was transferred to a 30 cc UV/light-block amber syringe (Nordson
EFD) and degassed by centrifugation at 2500 rpm for 10 min (Avanti J-25
I) to remove trapped air.

To increase the elastic modulus of the passive elastomer, it was blended
with a stiffer elastomer ink at ratios of 1:10 and 1:4 (stiffer ink:
passive ink). The stiffer ink was prepared from Ebecryl 8413 resin and
pentaerythritol tetraacrylate (1:1 by weight) with 10 wt.\% fumed silica
(CAB-O-SIL EH-5) and 4 wt.\% Irgacure I-651, mixed using the same
protocol described above.

\textbf{Rotational multimaterial 3D printing:} Printing was performed
following a previously reported protocol (56) with minor modifications.
Separate syringes containing the active and passive elastomer inks were
mounted onto a custom dual-channel nozzle fabricated using a D4K digital
light processing (DLP) printer from a rigid photopolymer resin
(Supplementary Fig. 1). The nozzle comprised two semi-circular internal
channels that merged at a 1 mm-diameter outlet to enable co-extrusion.
Each syringe was connected to an independent digital pressure controller
(PCD-100PSIG-D, Alicat Scientific) through pneumatic couplings, enabling
synchronized pressure control during print motion.

Filaments, printed with or without nozzle rotation, were deposited onto
cleaned glass substrates and subsequently UV-cured under an argon
atmosphere for 10 min per side. UV curing was performed using a UV lamp
(Omnicure Series 2000, 200 W, 320--500 nm; Lumen Dynamics Inc.) housed
in a UV box. Lattices were printed layer-by-layer on glass substrates
and UV-cured after each layer. Following completion of the print, the
entire lattice was UV-cured for an additional 10 min per side under
argon to ensure complete crosslinking. After curing, filaments and
lattices were carefully removed from the substrates using a fresh razor
blade.

\textbf{Thermal characterization:} Differential scanning calorimetry
(DSC; Discovery DSC 250, TA Instruments) was used to determine the
nematic-to-isotropic transition temperature ($T_\text{NI}$) of the
active liquid crystal elastomer ink. Samples ($\ge$5 mg) were sealed in
aluminum pans and subjected to a heating--cooling--reheating cycle from
room temperature to 200 $^\circ$C, cooling to -50 $^\circ$C, and reheating to 200 $^\circ$C
at 10 $^\circ$C min$^{-1}$. $T_\text{NI}$ was defined as the peak of the
melting endotherm obtained from the second heating cycle and determined
using the enthalpy-of-melt analysis tool in TA TRIOS software.
Measurements were performed in triplicate.

The glass transition temperature (T\textsubscript{g}) of the crosslinked
active and passive elastomers was measured using DSC under similar
conditions. Samples ($\ge$4 mg) were heated to 200 $^\circ$C, cooled to -90 $^\circ$C, and
reheated to 200 $^\circ$C. Heating was performed at 10 $^\circ$C min$^{-1}$ and cooling at
5 $^\circ$C min$^{-1}$. A slower cooling rate was used to reduce kinetic
undercooling and improve resolution of the glass-transition features.
T\textsubscript{g} was determined from the second heating cycle using
the glass-transition analysis tool in TA TRIOS software. All
measurements were conducted in triplicate.

\textbf{Actuation measurements:} Actuation of bulk materials and
architected multimaterial filaments and lattices was characterized by
immersing samples in a temperature-controlled silicone oil bath. Images
and videos were recorded using a uEye camera (Imaging Development
Systems Inc.), while temperature was monitored with a thermocouple
(Fluke T3000). The normalized lengths of bulk active and passive
elastomer samples were measured during heating from 25 $^\circ$C to 150 $^\circ$C.
Bilayer curvatures (active/passive), printed at varying speeds without
angular rotation, were measured during heating from 25 $^\circ$C to 175 $^\circ$C.

For architected filaments printed at varying angular rotation rates and
a constant translation velocity of 3 mm s$^{-1}$, the
end-to-end normalized length change was determined by imaging at 25 $^\circ$C,
heating to 175 $^\circ$C, and reimaging. Area changes in homogeneous
architected lattices were quantified during heating and cooling between
25 $^\circ$C and 175 $^\circ$C. Thermal cycling tests were conducted by alternating
samples between oil baths maintained at 25 $^\circ$C and 175 $^\circ$C.

Heterogeneous lattices were imaged using dual cameras (uEye for top view
and Canon Rebel EOS Ti2 for side view). For all samples, images were
captured at 25 $^\circ$C prior to heating to 175 $^\circ$C. All measurements were
performed on three independently prepared samples and analyzed using
ImageJ (version 1.53t). Cyclic actuation of Janus filaments printed
without rotation was evaluated over 100 cycles by alternating immersion
between oil baths at 25 $^\circ$C and 175 $^\circ$C, with imaging performed at 25 $^\circ$C
before each heating step.

\textbf{Rheological and mechanical characterization:} The rheological
properties of the active and passive elastomer inks were measured using
a rheometer (TA Instruments HR-20) equipped with a 20 mm steel Peltier
parallel-plate geometry and a 300 µm gap. After loading, samples were
heated to 125 $^\circ$C and held for 300 s to erase thermal history, then
cooled to 26 $^\circ$C and equilibrated for 500 s prior to measurement.
Viscosity was measured as a function of shear rate using a logarithmic
sweep from 0.01 s$^{-1}$ to 1000 s$^{-1}$. All measurements were performed on
three independently prepared samples.

Tensile properties of bulk elastomers and active-passive multimaterial
filaments were measured using the same instrument equipped with tensile
grips. Printed filaments (\textasciitilde600 µm diameter, 30 mm length)
were strained at 100 $\mu$m s$^{-1}$ until failure. Measurements were conducted
on three samples per condition.

Bilayer adhesion tests were performed using the hybrid rheometer
equipped with tensile grips. Samples were prepared by injection molding
the active and passive elastomer inks into rectangular specimens
incorporating a 1 mm-wide interfacial gap to provide an engineered
delaminated region for gripping. The two elastomers were crosslinked
simultaneously. During testing, the active elastomer region was clamped
in the lower grip and the passive elastomer free end in the upper grip,
followed by tensile loading at 100 µm s$^{-1}$ until failure. All adhesion
measurements were performed in triplicate.

\textbf{Wide-angle X-ray scattering (WAXS):} Laboratory X-ray
measurements were performed in transmission mode using a Xeuss 3.0
beamline (Xenocs Inc.). Scattering patterns were collected at 8.04 keV
with a beam spot size of \textasciitilde1.4 mm and an exposure time of
300 s using a Pilatus 300K detector (Dectris). A single stitched
scattering pattern was constructed from measurements acquired at
multiple detector positions to eliminate detector gaps. Background
subtraction was performed using empty-air measurements for pure active
LCE samples and pure passive elastomer measurements for active-passive
composites. Sample-to-detector distance calibration was carried out
using a LaB$_6$ diffraction standard.

Scalar orientational order parameters,
\(S = \langle P_{2}(\cos\chi)\rangle\), were determined by integrating
the mesogen scattering peak intensity over the range
\(q = 0.65\text{-}2.5\ {Å}^{- 1}\) as a function of azimuthal angle
and fitting the resulting distribution using the Kratky method (75).
Data reduction and stitching were performed using custom Python scripts
based on pyFAI (76).

The spatial phase behavior of printed fibers was further characterized
by transmission WAXS at the Soft Matter Interfaces beamline (12-ID SMI)
at the National Synchrotron Light Source II (NSLS-II), Brookhaven
National Laboratory (77). Scattering patterns were recorded using a
Pilatus 900K-W detector with an incident X-ray energy of 16.1 keV. The
beam dimensions (full width at half maximum) were 25 µm (horizontal) ×
2.5 µm (vertical), and the beam was aligned along the x-axis,
perpendicular to the nematic director. Measurements were conducted under
ultra-high vacuum at room temperature. Order parameters were calculated
as described above.

\textbf{Microscopy:} Scanning electron microscopy (Zeiss Gemini 360
FE-SEM) (SEM) was used to image cross-sections of the filaments prepared
through nitrogen freeze-fracture. Liquid nitrogen was placed in a
Styrofoam container and held under vacuum for approximately 5 minutes to
bring its temperature below the glass transition temperature of both
elastomers. The fiber samples were then immediately submerged in liquid
nitrogen. After 2--5 minutes, the fibers were fractured by snapping them
in half with two self-closing tweezers. The fractured surfaces were
mounted onto 90° stubs (Ted Pella) using conductive carbon tape. A 10 nm
Pt/Pd conductive coating was subsequently sputtered onto the mounted
samples using an EMS150T S sputter coater. Bright field microscopy (Axio
Observer, Zeiss) was used to image filaments. Images were analyzed using
ImageJ software (version 1.53t).

\textbf{Active filter:} The filter assembly consisted of a heterogeneous
lattice designed to contract along its perimeter while expanding in the
center, thereby minimizing out-of-plane deformation under edge
constraints. The lattice was mechanically secured to a laser-cut acrylic
frame using 3 mm screws to attach the lattice to the top plate and 16 mm
screws to fasten the top plate to the bottom support plate. The
assembled structure was immersed in a silicone oil bath and imaged from
above using a uEye camera while temperature was monitored with a
thermocouple. To demonstrate filtering functionality, grinding media
spheres (5 mm diameter, 450 mg; Inframat Advanced Materials) and M4
bolts (7 mm end-to-end length, 650 mg) were used as representative
objects for the closed and open filter states.

\textbf{Pick and place:} The gripper consisted of a contracting lattice
attached to a laser-cut acrylic square base using M3 screws and bolts.
The PDMS base was fabricated by replica molding from a rigid positive
master printed using a D4K DLP printer. The printed master (120 mm × 75
mm × 6 mm) contained raised square platforms (19 mm × 19 mm × 3 mm),
cylindrical posts (6 mm diameter, 3 mm height) defining the placement
slots, and an array of raised alignment markers. A PDMS prepolymer
mixture (Sylgard 184), dyed with red Silc Pig pigment, was poured over
the printed master and cured at 75 $^\circ$C for 12 h. After curing, the PDMS
base was peeled from the master, yielding a negative replica with
recessed square features, cylindrical cavities, and alignment dimples.
Acrylic rods were initially positioned on the PDMS surface using the
alignment dimples as fiducial markers. The gripper was then placed over
the rods, and the assembly was heated to 175 $^\circ$C in a silicone oil bath
(temperature monitored by thermocouple) to contract the lattice and
grasp the rods. The gripper carrying the rods was transferred to the
target region containing the recessed cavities, and the bath was cooled
to 50 $^\circ$C to release the rods as the lattice expanded. Video was recorded
from the top view using a uEye camera.

\textbf{Discrete Elastic Rod (DER) Model:} We employed the discrete
elastic rod (DER) model to efficiently compute the minimal elastic
energy configurations (58,78). The filament's centerline is discretized
into a set of \(N\) vertices \(\mathbf{r}_{s}\) connected by edges
\(\mathbf{e}_{s} = \mathbf{r}_{s + 1} - \mathbf{r}_{s}\), where the
previously continuous arc length parameter now \(s\) becomes an integer
index along the filament's centerline. Discrete curvature and twist are
defined at each interior vertex \(s\).

With the material frame
\{\(\mathbf{d}_{1,s},\mathbf{\ d}_{2,s},\mathbf{d}_{3,s}\)\} defined at
each interior vertex \(s\), the curvature binormal (which captures
bending) is then computed as:

\[\left( \mathbf{k}_{b} \right)_{s} = \frac{2\mathbf{e}_{s - 1} \times \mathbf{e}_{s}}{\left| \mathbf{e}_{s - 1} \right|\left| \mathbf{e}_{s} \right| + \mathbf{e}_{s - 1} \cdot \mathbf{e}_{s}}\]

The components \(k_{1,s}\) and \(k_{2,s}\) are the projections of
\(\left( \mathbf{k}_{b} \right)_{s}\) onto the local basis vectors
\(\mathbf{d}_{1,s}\) and \(\mathbf{d}_{2,s}\), while the discrete twist
\(\tau_{s}\) measures rotation of the material frame between adjacent
edges. The continuous elastic energy integral is approximated by a sum
over the vertices. The energy at each vertex \(s\) is weighted by the
inverse of the Voronoi length
\({\bar{l}}_{s} = \left( \left| \mathbf{e}_{s - 1} \right| + \left| \mathbf{e}_{s} \right| \right)/2\).
The total discrete elastic energy is then given by:\\
\[E_{\text{discrete}} = \sum_{s = 1}^{N - 1}\frac{1}{2{\bar{l}}_{s}}\left\lbrack B\left( \left( \kappa_{1,s} - {\bar{\kappa}}_{1,s} \right)^{2} + \left( \kappa_{2,s} - {\bar{\kappa}}_{2,s} \right)^{2} \right) + C\left( \tau_{s} - {\bar{\tau}}_{s} \right)^{2} \right\rbrack\]

where \(B\) and \(C\) are the bending and torsional rigidities,
\({\bar{l}}_{s}\) is the Voronoi length and \({\bar{\kappa}}_{1,s}\),
\({\bar{\kappa}}_{2,s}\), and \({\bar{\tau}}_{s}\) are the natural
curvatures and twist, respectively. The equilibrium configuration
corresponds to the set of vertex positions \(\{\mathbf{r}_{s}\}\) that
minimize \(E_{\text{discrete}}\), found using a standard gradient
descent algorithm implemented in the open-source package disMech (78)~.

\textbf{Numerical simulation}: We model the architected lattice as a
discrete graph consisting of 25 branch nodes---corresponding to
junctions where sinusoidal filaments meet---and 40 connecting edges.
Each edge is discretized further into 10 ghost nodes to resolve the
sinusoidal geometry, resulting in a total of 425 nodes (Supplementary
Fig. 13A,B).

\emph{Parametrizing the shape-morphing lattice in numerical
simulation}\emph{s}: First, we consider a simple square grid with \(n\)
points on each side. The grid has a total of \(N = n^{2}\) points, which
we call vertices. We assign each one a unique number (index) from \(0\)
to \(N - 1\). To find the index \(k\) for a vertex at row \(i\) and
column \(j\), we use \emph{row-major order}: \(k = (n \times i) + j\).
If the distance between any two adjacent points is \(l\), the spatial
coordinate \(\mathbf{v}_{k}\) for the vertex at \((i,j)\) is given by:
\(\mathbf{v}_{i,j} = (l \cdot i,l \cdot j)\).

\emph{Creating wavy edges:} Next, we replace the straight lines (edges)
connecting the vertices with curves. To do this, we add a series of new
points, called \emph{ghost nodes}, along the path of each original edge.
These ghost nodes are not placed in a straight line; they are offset to
form a sine wave, making the edge appear wavy.

\emph{The formula for ghost nodes}: To calculate the exact position of
each ghost node along an edge, we use the following method. Consider a
straight edge from a starting vertex \(\mathbf{v}_{\text{start}}\) to an
ending vertex \(\mathbf{v}_{\text{end}}\). To place \(n_{g}\) ghost
nodes along it, we define: (1) the unit vector \(\mathbf{T}\) pointing
from \(\mathbf{v}_{\text{start}}\) to \(\mathbf{v}_{\text{end}}\) which
is the direction of the straight edge and (2) the unit vector
\(\mathbf{N}\) perpendicular to \(\mathbf{T}\) which is the direction of
the wave's oscillation, and (3) the amplitude, or maximum height, of the
wave, \(A\). The position of the \(s\)-th ghost node (where \(s\) is an
integer from \(1\) to \(n_{g}\)) is a combination of moving along the
straight edge and then adding a perpendicular offset for the wave:

\[\mathbf{v}_{\text{ghost}}(s) = \underset{\text{Position along straight line}}{\underbrace{\mathbf{v}_{\text{start}} + \mathbf{T}\left( \frac{s \cdot l}{n_{g} + 1} \right)}} + \underset{\text{Perpendicular wave offset}}{\underbrace{\mathbf{N} \cdot A\sin\left( \frac{2\pi s}{n_{g} + 1} \right)}}\]

\emph{Gradual increase in natural curvature amplitude:} We simulate the
continuous transformation from a flat sheet to the final 3D object by
incrementally increasing the constant \(x\) from 0 to 0.75, so that
\(c_{s} = 1 + x\) ranges from 1 to 1.75 (for straightening edges) and
\(c_{t} = 1 - x\) from 1 to 0.25 (for curling edges). At each step, the
structure's stable equilibrium shape is found by minimizing its total
elastic energy using a simple gradient descent method (78).

\textbf{Acknowledgments}

The authors gratefully acknowledge support from the National Science
Foundation through the Harvard MRSEC (DMR-2011754) and the ARO MURI
program (W911NF-17-1-03; W911NF-22-1-0219). M. Abdelrahman was supported
through an MPS-ASCEND Postdoctoral Fellowship. J. Wilt was supported
through the NSF-GRFP Fellowship. L.M. was supported partially by the
Simons Foundation and the Henri Seydoux Fund. This work made use of the
Shared Experimental Facilities supported in part by the Harvard MRSEC,
and this work was performed in part at the Harvard University Center for
Nanoscale Systems (CNS); a member of the National Nanotechnology
Coordinated Infrastructure Network (NNCI), which is supported by the
National Science Foundation under NSF award no. ECCS-2025158. This
research used SMI 12-ID of the National Synchrotron Light Source II, a
US DOE Office of Science User Facility operated for the DOE Office of
Science by Brookhaven National Laboratory under contract number
DE-SC0012704. We thank Dr. Patryk Wasik for assisting in data collection
at NSLSII.

\textbf{References}

1. V. P. Patil, J. D. Sandt, M. Kolle, J. Dunkel, Topological mechanics
of knots and tangles. \emph{Science} 367, 71--75 (2020).

2. K. A. Dill, J. L. MacCallum, The protein-folding problem, 50 years
on. \emph{Science} 338, 1042--1046 (2012).

3. S. J. Gerbode, J. R. Puzey, A. G. McCormick, L. Mahadevan, How the
cucumber tendril coils and overwinds. \emph{Science} 337, 1087--1091
(2012).

4. C. Darwin, The movements and habits of climbing plants (John Murray,
1875).

5. M. S. Sousa-Baena, N. R. Sinha, J. Hernandes-Lopes, L. G. Lohmann,
Convergent evolution and the diverse ontogenetic origins of tendrils in
angiosperms. \emph{Front. Plant Sci.} 9, 403 (2018).

6. J. N. Richter, B. Hochner, M. J. Kuba, Octopus arm movements under
constrained conditions: adaptation, modification and plasticity of motor
primitives. \emph{J. Exp. Biol.} 218, 1069--1076 (2015).

7. L. Purkart, et al., Trigeminal ganglion and sensory nerves suggest
tactile specialization of elephants. \emph{Curr. Biol.} 32, 904-910.e3
(2022).

8. L. L. Longren, et al., Dense reconstruction of elephant trunk
musculature. \emph{Curr. Biol.} 33, 4713-4720.e3 (2023).

9. P. Dagenais, S. Hensman, V. Haechler, M. C. Milinkovitch, Elephants
evolved strategies reducing the biomechanical complexity of their trunk.
\emph{Curr. Biol.} 31, 4727-4737.e4 (2021).

10. C. G. Meloche, J. P. Knox, K. C. Vaughn, A cortical band of
gelatinous fibers causes the coiling of redvine tendrils: a model based
upon cytochemical and immunocytochemical studies. \emph{Planta} 225,
485--498 (2007).

11. A. J. Bowling, K. C. Vaughn, Gelatinous fibers are widespread in
coiling tendrils and twining vines. \emph{Am. J. Bot.} 96, 719--727
(2009).

12. A. K. Schulz, et al., Skin wrinkles and folds enable asymmetric
stretch in the elephant trunk. \emph{Proc. Natl. Acad. Sci. U.S.A.} 119,
e2122563119 (2022).

13. M. K. Abdelrahman, et al., Material assembly from collective action
of shape-changing polymers. \emph{Nat. Mater.} 23, 281--289 (2024).

14. J. K. Wilt, N. M. Larson, J. A. Lewis, Rotational multimaterial 3D
printing of soft robotic matter with embedded asymmetrical pneumatics.
\emph{Adv. Mater.} \textbf{38}, e10141 (2026).

15. S. Zou, et al., A retrofit sensing strategy for soft fluidic robots.
\emph{Nat. Commun.} 15, 539 (2024).

16. K. Becker, et al., Active entanglement enables stochastic,
topological grasping. \emph{Proc. Natl. Acad. Sci. U.S.A.} 119,
e2209819119 (2022).

17. M. Schaffner, et al., 3D printing of robotic soft actuators with
programmable bioinspired architectures. \emph{Nat. Commun.} 9, 878
(2018).

18. H. Kim, et al., Inherently integrated microfiber-based flexible
proprioceptive sensor for feedback-controlled soft actuators. \emph{npj
Flexible Electronics} 8, 15 (2024).

19. X. Wan, \emph{et al.}, Multimaterial shape memory polymer fibers for
advanced drug release applications. \emph{Adv. Fiber Mater.} \textbf{7},
1576--1589 (2025).

20. B. J. Woodington, et al., Electronics with shape actuation for
minimally invasive spinal cord stimulation. \emph{Sci. Adv.} 7, eabg7833
(2021).

21. A. Lendlein, R. Langer, Biodegradable, elastic shape-memory polymers
for potential biomedical applications. \emph{Science} 296, 1673--1676
(2002).

22. E. Lee, D. Kim, H. Kim, J. Yoon, Photothermally driven fast
responding photo-actuators fabricated with comb-type hydrogels and
magnetite nanoparticles. \emph{Sci. Rep.} 5, 15124 (2015).

23. D. J. Roach, et al., 4D printed multifunctional composites with
cooling‐rate mediated tunable shape morphing. \emph{Adv. Funct. Mater.}
32 (2022).

24. X. Li, X. Cai, Y. Gao, M. J. Serpe, Reversible bidirectional bending
of hydrogel-based bilayer actuators. \emph{J. Mater. Chem. B} 5,
2804--2812 (2017).

25. A. S. Gladman, E. A. Matsumoto, R. G. Nuzzo, L. Mahadevan, J. A.
Lewis, Biomimetic 4D printing. \emph{Nat. Mater.} 15, 413--418 (2015).

26. J. Bae, J.-H. Na, C. D. Santangelo, R. C. Hayward, Edge-defined
metric buckling of temperature-responsive hydrogel ribbons and rings.
\emph{Polymer} 55, 5908--5914 (2014).

27. W. M. van Rees, E. A. Matsumoto, A. S. Gladman, J. A. Lewis, L.
Mahadevan, Mechanics of biomimetic 4D printed structures. \emph{Soft
Matter} 14, 8771--8779 (2018).

28. W. M. van Rees, E. Vouga, L. Mahadevan, Growth patterns for
shape-shifting elastic bilayers. \emph{Proc. Natl. Acad. Sci. U.S.A.}
114, 11597--11602 (2017).

29. J. W. Boley, et al., Shape-shifting structured lattices via
multimaterial 4D printing. \emph{Proc. Natl. Acad. Sci. U.S.A.} 116,
20856--20862 (2019).

30. E. M. P. Cosserat, F. Cosserat, \emph{Théorie des corps déformables}
(A. Hermann et fils, Paris, 1909).

31. S. S. Antman, \emph{Nonlinear Problems of Elasticity} (Springer, New
York, 2005).

32. Y. Xia, G. M. Whitesides, Soft Lithography. \emph{Angew. Chem. Int.
Ed.} 37, 550--575 (1998).

33. R. F. Shepherd, et al., Multigait soft robot. \emph{Proc. Natl.
Acad. Sci. U.S.A.} 108, 20400--3 (2011).

34. T. J. Jones, E. Jambon-Puillet, J. Marthelot, P.-T. Brun, Bubble
casting soft robotics. \emph{Nature} 599, 229--233 (2021).

35. Z. Zhang, et al., Soft and lightweight fabric enables powerful and
high-range pneumatic actuation. \emph{Sci. Adv.} 9, eadg1203 (2023).

36. F. Connolly, C. J. Walsh, K. Bertoldi, Automatic design of
fiber-reinforced soft actuators for trajectory matching. \emph{Proc.
Natl. Acad. Sci. U.S.A.} 114, 51--56 (2017).

37. M. Warner, E. M. Terentjev, \emph{Liquid Crystal Elastomers} (Oxford
University Press, Oxford, 2007).

38. Q. He, et al., Electrospun liquid crystal elastomer microfiber
actuator. \emph{Sci. Robot.} 6 (2021).

39. C. M. Yakacki, \emph{et al.}, Tailorable and programmable
liquid-crystalline elastomers using a two-stage thiol--acrylate
reaction. \emph{RSC Adv.} \textbf{5}, 18997--19001 (2015).

40. T. H. Ware, M. E. McConney, J. J. Wie, V. P. Tondiglia, T. J. White,
Voxelated liquid crystal elastomers. \emph{Science} 347, 982--984
(2015).

41. J. T. Waters, et al., Twist again: Dynamically and reversibly
controllable chirality in liquid crystalline elastomer microposts.
\emph{Sci. Adv.} 6, eaay5349 (2020).

42. H. Yang, et al., Micron-sized main-chain liquid crystalline
elastomer actuators with ultralarge amplitude contractions. \emph{J. Am.
Chem. Soc.} 131, 15000--15004 (2009).

43. S. Li, M. Aizenberg, M. M. Lerch, J. Aizenberg, Programming
deformations of 3D microstructures: opportunities enabled by magnetic
alignment of liquid crystalline elastomers. \emph{Acc. Mater. Res.} 4,
1008--1019 (2023).

44. A. Kotikian, R. L. Truby, J. W. Boley, T. J. White, J. A. Lewis, 3D
printing of liquid crystal elastomeric actuators with spatially
programed nematic order. \emph{Adv. Mater.} 30 (2018).

45. H. Zeng, O. M. Wani, P. Wasylczyk, R. Kaczmarek, A. Priimagi,
Self‐regulating iris based on light‐actuated liquid crystal elastomer.
\emph{Adv. Mater.} 29 (2017).

46. D. J. Roach, et al., Long liquid crystal elastomer fibers with large
reversible actuation strains for smart textiles and artificial muscles.
\emph{ACS Appl. Mater. Interfaces} 11, 19514--19521 (2019).

47. P. E. S. Silva, et al., Active textile fabrics from weaving liquid
crystalline elastomer filaments. \emph{Adv. Mater.} 35, e2210689 (2023).

48. Y. Geng, R. Kizhakidathazhath, J. P. F. Lagerwall, Robust
cholesteric liquid crystal elastomer fibres for mechanochromic textiles.
\emph{Nat. Mater.} 21, 1441--1447 (2022).

49. J.-H. Lee, \emph{et al.}, Redefining the limits of actuating fibers
via mesophase control: from contraction to elongation. \emph{Sci. Adv.}
\textbf{11}, eadt7613 (2025).

50. A. Kotikian, et al., Innervated, self‐sensing liquid crystal
elastomer actuators with closed loop control. \emph{Adv. Mater.} 33,
e2101814 (2021).

51. Y. Yang, et al., Near‐infrared light‐driven MXene/liquid crystal
elastomer bimorph membranes for closed‐loop controlled self‐sensing
bionic robots. \emph{Adv. Sci.} 11, 2307862 (2023).

52. H. Liu, et al., Shape-programmable, deformation-locking, and
self-sensing artificial muscle based on liquid crystal elastomer and
low-melting point alloy. \emph{Sci. Adv.} 8, eabn5722 (2022).

53. S. J. D. Lugger, et al., Melt‐extruded thermoplastic liquid crystal
elastomer rotating fiber actuators. \emph{Adv. Funct. Mater.} 33 (2023).

54. W. Seo, et al., Azobenzene‐functionalized semicrystalline liquid
crystal elastomer springs for underwater soft robotic actuators.
\emph{Small} e2406493 (2024).

55. Z. Hu, Y. Li, T. Zhao, J. Lv, Self-winding liquid crystal elastomer
fiber actuators with high degree of freedom and tunable actuation.
\emph{Appl. Mater. Today} 27, 101449 (2022).

56. N. M. Larson, et al., Rotational multimaterial printing of filaments
with subvoxel control. \emph{Nature} 613, 682--688 (2023).

57. L. T. de Haan, A. P. H. J. Schenning, D. J. Broer, Programmed
morphing of liquid crystal networks. \emph{Polymer} 55, 5885--5896
(2014).

58. M. Bergou, M. Wardetzky, S. Robinson, B. Audoly, E. Grinspun,
Discrete elastic rods. \emph{ACM SIGGRAPH} 2008 Pap. 1--12 (2008).

59. M. Gazzola, L. H. Dudte, A. G. McCormick, L. Mahadevan, Forward and
inverse problems in the mechanics of soft filaments. \emph{R. Soc. Open
Sci.} 5, 171628 (2018).

60. S. Timoshenko, Analysis of bi-metal thermostats. \emph{J. Opt. Soc.
Am.} 11, 233 (1925).

61. R. Telles, et al., Spatially programmed alignment and actuation in
printed liquid crystal elastomers. \emph{Proc. Natl. Acad. Sci. U.S.A.}
122, e2414960122 (2025).

62. L. Ren, et al., Programming shape-morphing behavior of liquid
crystal elastomers via parameter-encoded 4D printing. \emph{ACS Appl.
Mater. Interfaces} 12, 15562--15572 (2020).

63. J. R. Raney, et al., Rotational 3D printing of damage-tolerant
composites with programmable mechanics. \emph{Proc. Natl. Acad. Sci.
U.S.A.} 115, 1198--1203 (2018).

64. A. Giudici, J. S. Biggins, Large deformation analysis of spontaneous
twist and contraction in nematic elastomer fibers with helical director.
\emph{J. Appl. Phys.} 129, 154701 (2021).

65. T. C. T. Michaels, E. Memet, L. Mahadevan, Mechanical basis for
fibrillar bundle morphology. \emph{Soft Matter} 16, 9306--9318 (2020).

66. A. T. Liu, et al., Colloidal robotics. \emph{Nat. Mater.} 22,
1453--1462 (2023).

67. Z. Liu, et al., 3D-printed low-voltage-driven ciliary hydrogel
microactuators. \emph{Nature} 649, 885--893 (2026).

68. D. W. Thompson, \emph{On Growth and Form} (Cambridge University
Press, 1917).

69. E. Coen, D. J. Cosgrove, The mechanics of plant morphogenesis.
\emph{Science} 379, eade8055 (2023).

70. W. Neveu, I. Puhachov, B. Thomaszewski, M. Bessmeltsev,
Stability-aware simplification of curve networks. \emph{Spec. Interest
Group Comput. Graph. Interact. Tech. Conf. Proc.} 1--9 (2022).

71. D. Duffy, J. M. McCracken, T. S. Hebner, T. J. White, J. S. Biggins,
Lifting, loading, and buckling in conical shells. \emph{Phys. Rev.
Lett.} 131, 148202 (2023).

72. J. Kim, J. A. Hanna, M. Byun, C. D. Santangelo, R. C. Hayward,
Designing responsive buckled surfaces by halftone gel lithography.
\emph{Science} 335, 1201--1205 (2012).

73. H. Zhang, X. Guo, J. Wu, D. Fang, Y. Zhang, Soft mechanical
metamaterials with unusual swelling behavior and tunable stress-strain
curves. \emph{Sci. Adv.} 4, eaar8535 (2018).

74. X. Ni, et al., 2D mechanical metamaterials with widely tunable
unusual modes of thermal expansion. \emph{Adv. Mater.} 31, e1905405
(2019).

75. M. T. Sims, L. C. Abbott, R. M. Richardson, J. W. Goodby, J. N.
Moore, Considerations in the determination of orientational order
parameters from X-ray scattering experiments. \emph{Liq. Cryst.} 46,
11--24 (2019).

76. J. Kieffer, D. Karkoulis, PyFAI, a versatile library for azimuthal
regrouping. \emph{J. Phys.: Conf. Ser.} 425, 202012 (2013).

77. M. Zhernenkov, N. Canestrari, O. Chubar, E. DiMasi, Soft matter
interfaces beamline at NSLS-II: geometrical ray-tracing vs. wavefront
propagation simulations. \emph{Adv. Comput. Methods X-Ray Opt. III}
92090G-92090G--9 (2014).

78. A. Choi, R. Jing, A. P. Sabelhaus, M. K. Jawed, DisMech: A Discrete
Differential Geometry-Based Physical Simulator for Soft Robots and
Structures. \emph{IEEE Robot. Autom. Lett.} 9, 3483--3490 (2024).

\clearpage
\textbf{Figures}

\includegraphics[width=6in,height=3.5858in]{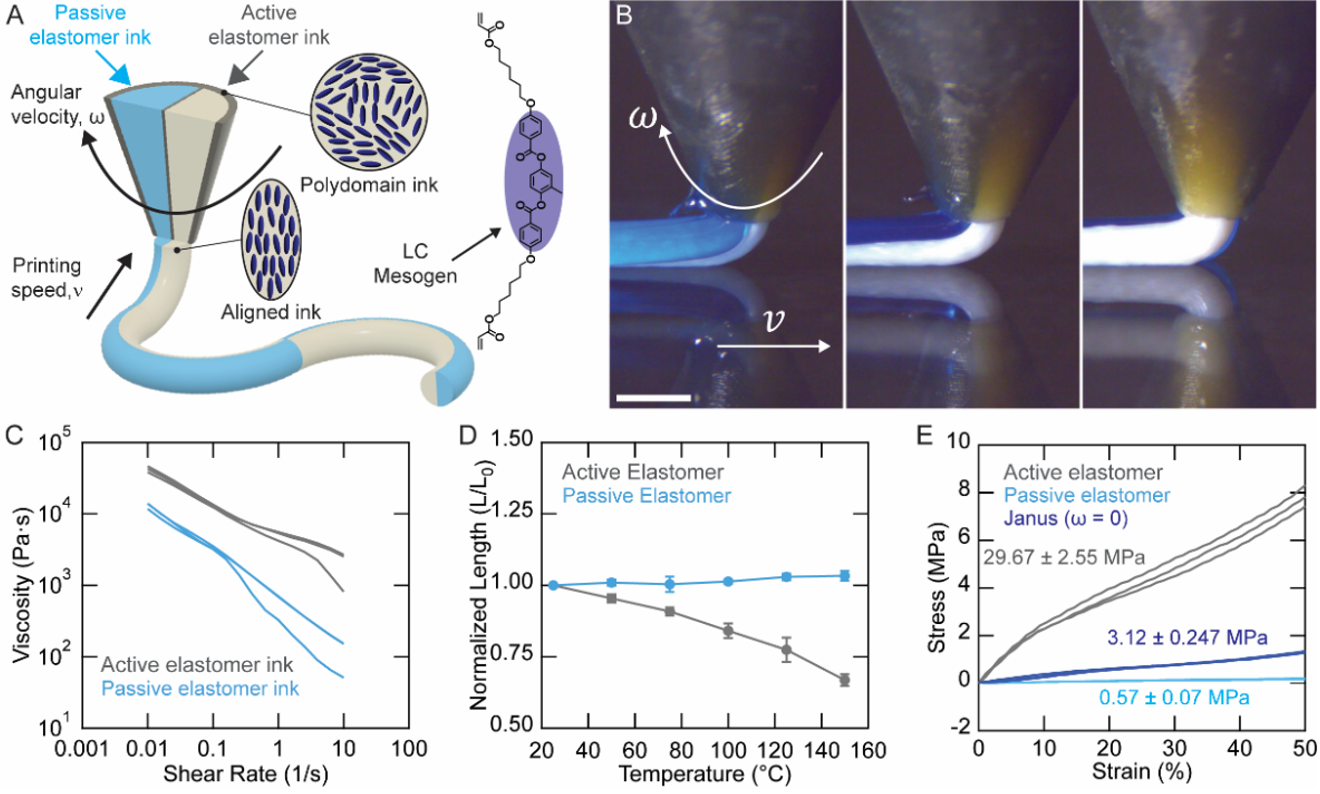}

\textbf{Figure 1. Rotational multimaterial 3D printing of active-passive
filaments. (A)} Schematic of co-extrusion of active and passive
elastomer inks through a customized 1 mm-diameter nozzle. \textbf{(B)}
Time-lapse images of ink co-extrusion during rotational printing. Scale
bar: 1 mm. \textbf{(C)} Apparent viscosity as a function of shear rate
for the active LCE (grey) and passive acrylate (light blue) elastomer
inks. \textbf{(D)} Normalized length as a function of temperature for
pure active LCE (grey) and passive acrylate (light blue) filaments. Data
are shown as mean $\pm$ s.d. (n = 3). \textbf{(E)} Uniaxial tensile
stress-strain curves for an active LCE filament (grey), a passive
acrylate filament (light blue), and an active-passive Janus filament
printed without rotation (navy blue). All measurements were performed on
three independently prepared batches (n = 3).

\includegraphics[width=5.53116in,height=6.32174in]{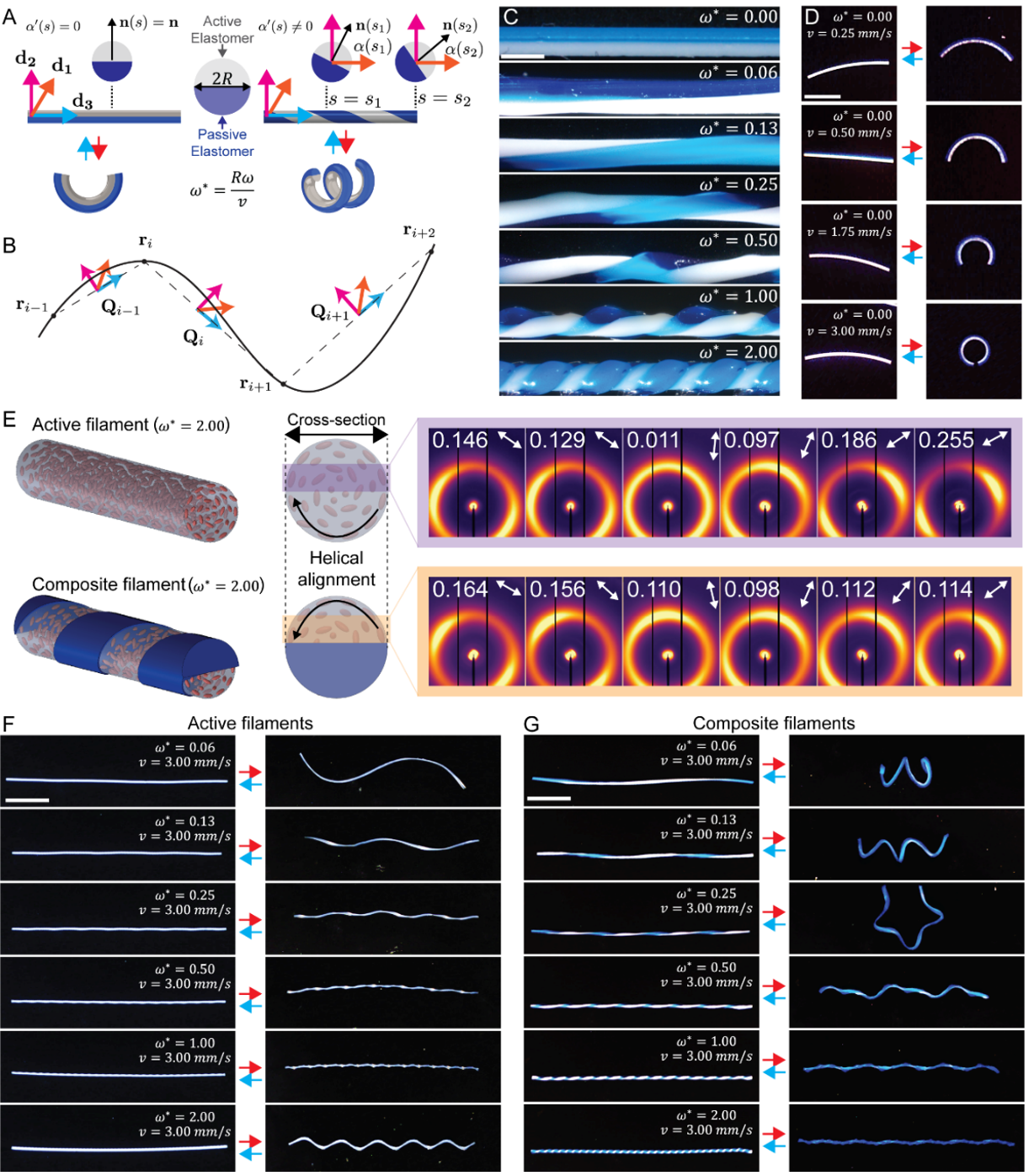}

\textbf{Figure 2. Programmable shape morphing of active-passive
filaments. (A)} Schematic of the material frame and interfacial normal
vector \(n(s)\). \textbf{(B)} Material distribution
\(\{ d_{1},\ d_{2},\ d_{3}\}\) at vertices \(i - 1\), \(i\), \(i + 1\).
\textbf{(C)} Optical images of architected filaments printed at varying
\(\omega^{*}\) with constant \(\nu = 3\) mm s$^{-1}$. Scale
bar: 1 mm. \textbf{(D)} Filaments printed at\(\ \omega^{*} = 0\) with
varying print speeds, shown in the initial state and heated states.
Scale bar: 5 mm. \textbf{(E)} Schematic of rotationally printed active
LCE and composite filaments, highlighting the helical mesogen
orientation imparted by rotational printing. Corresponding 2D WAXS
patterns measured across the filament cross-sections reveal the spatial
variation of the nematic director orientation. Numbers denote the scalar
order parameter (\(S\)), and arrows indicate the azimuthal angle of
maximum intensity ($X_\text{max}$). Both filaments were printed at
\(\omega^{*}\) = 2 and \(\nu = 3\) mm s$^{-1}$.
\textbf{(F)} Pure active filaments printed at varying \(\omega^{*}\)
with constant \(\nu = 3\ \) mm s$^{-1}$. Scale bar: 10 mm.
\textbf{(G)} Architected composite filaments printed at varying
\(\omega^{*}\) with constant \(\nu = 3\ \) mm s$^{-1}$.
Scale bar: 10 mm.

\includegraphics[width=6in,height=6.11334in]{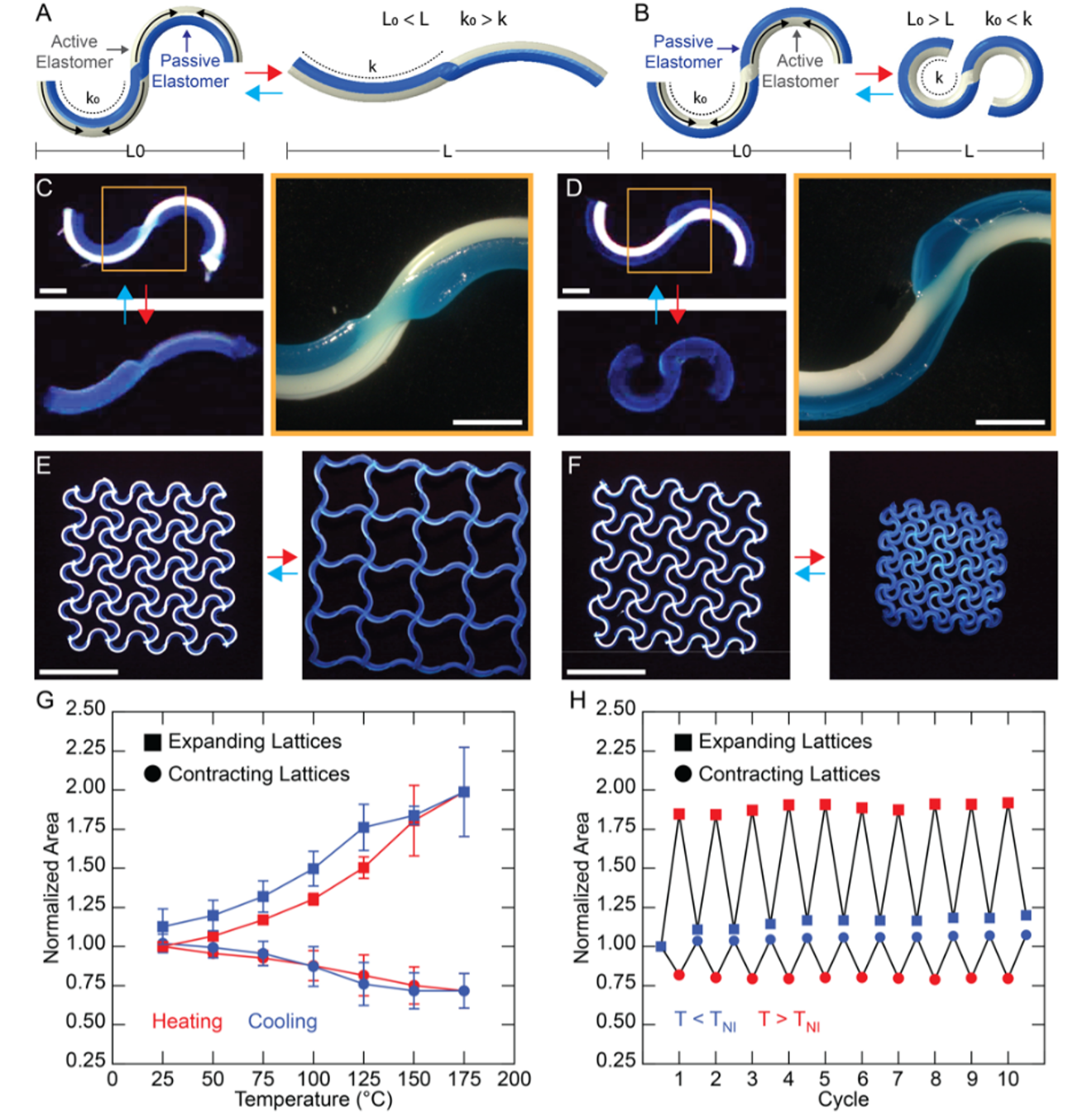}

\textbf{Figure 3. Active-passive lattices with homogeneous shape
morphing. (A)} Schematic of architected filaments with the active
elastomer printed on the outer edge, resulting in an increase in
filament end-to-end length upon heating. \textbf{(B)} Schematic of
architected filaments with the passive elastomer printed on the outer
edge, resulting in a reduction in filament end-to-end length upon
heating. \textbf{(C)} Expanding filament below and above
$T_\text{NI}$. Scale bar: 1 mm for both images. \textbf{(D)}
Contracting filament below and above $T_\text{NI}$. Scale bar: 1
mm for both images. \textbf{(E)} Three-layer lattice composed of
expanding filaments, exhibiting global lattice expansion upon heating.
Scale bar: 25 mm. \textbf{(F)} Three-layer lattice composed of
contracting filaments, exhibiting global lattice contraction upon
heating. Scale bar: 25 mm. \textbf{(G)} Normalized area as a function of
temperature for expanding lattices (squares) and contracting lattices
(circles) during heating (red lines) and cooling (blue lines), showing
the magnitude of areal changes. Data are shown as mean $\pm$ s.d. (n = 3).
\textbf{(H)} Normalized area as a function of cycle number for expanding
lattices (squares) and contracting lattices (circles) during cyclic
heating to 150 $^\circ$C (red points) and cooling to 25 $^\circ$C (blue points),
demonstrating reversibility.

\includegraphics[width=6in,height=3.3659in]{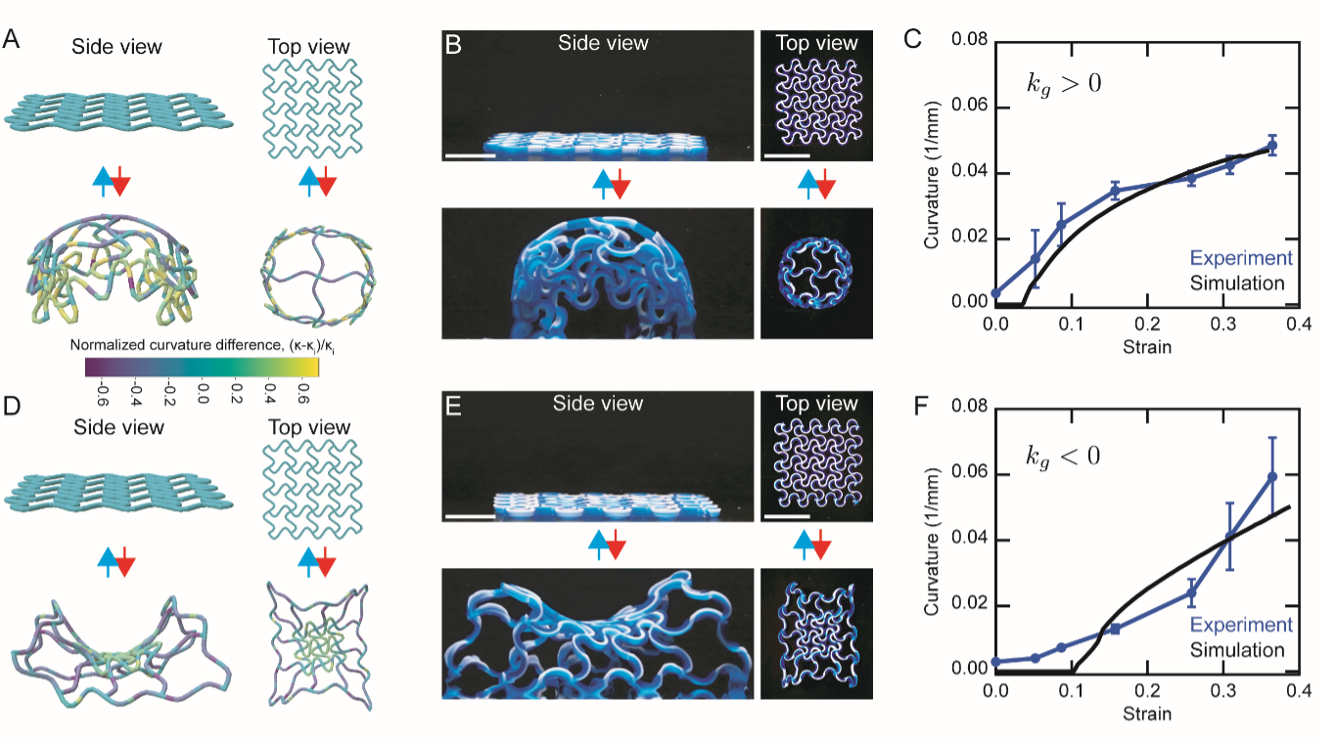}

\textbf{Figure 4. Active-passive lattices with heterogeneous shape
morphing. (A)} Simulated side and top views of a heterogeneous lattice
with expanding filaments in the central unit cells and contracting
filaments in the outer unit cells, shown in the initial flat state and
after heating. The curvature change relative to the initial state is
color-coded as \((\kappa - \kappa_{i})/\kappa_{i}\). \textbf{(B)}
Corresponding experimental side and top views before and after heating.
Scale bars: 10 mm (side view) and 25 mm (top view). \textbf{(C)}
Evolution of the characteristic curvature (\(\sqrt{|k_{1}k_{2}|}\)) as a
function of strain for lattices morphing into positive gaussian
curvature (\(\kappa_{g}\) \textgreater{} 0). Simulation (black) and
experiment (blue) are shown as mean $\pm$ s.d. (n = 3). \textbf{(D)}
Simulated side and top views of a heterogeneous lattice with contracting
filaments in the central unit cells and expanding filaments in the outer
unit cells, shown in the initial flat state and after heating.
\textbf{(E)} Corresponding experimental side and top views before and
after heating. Scale bar: 10 mm (side view) and 25 mm (top view).
\textbf{(F)} Evolution of the characteristic curvature
(\(\sqrt{|k_{1}k_{2}|}\)) as a function of strain for lattices morphing
into negative gaussian curvature (\(\kappa_{g}\) \textless{} 0).
Simulation (black) and experiment (blue) are shown as mean $\pm$ s.d. (n =
3).

\includegraphics[width=6in,height=5.96102in]{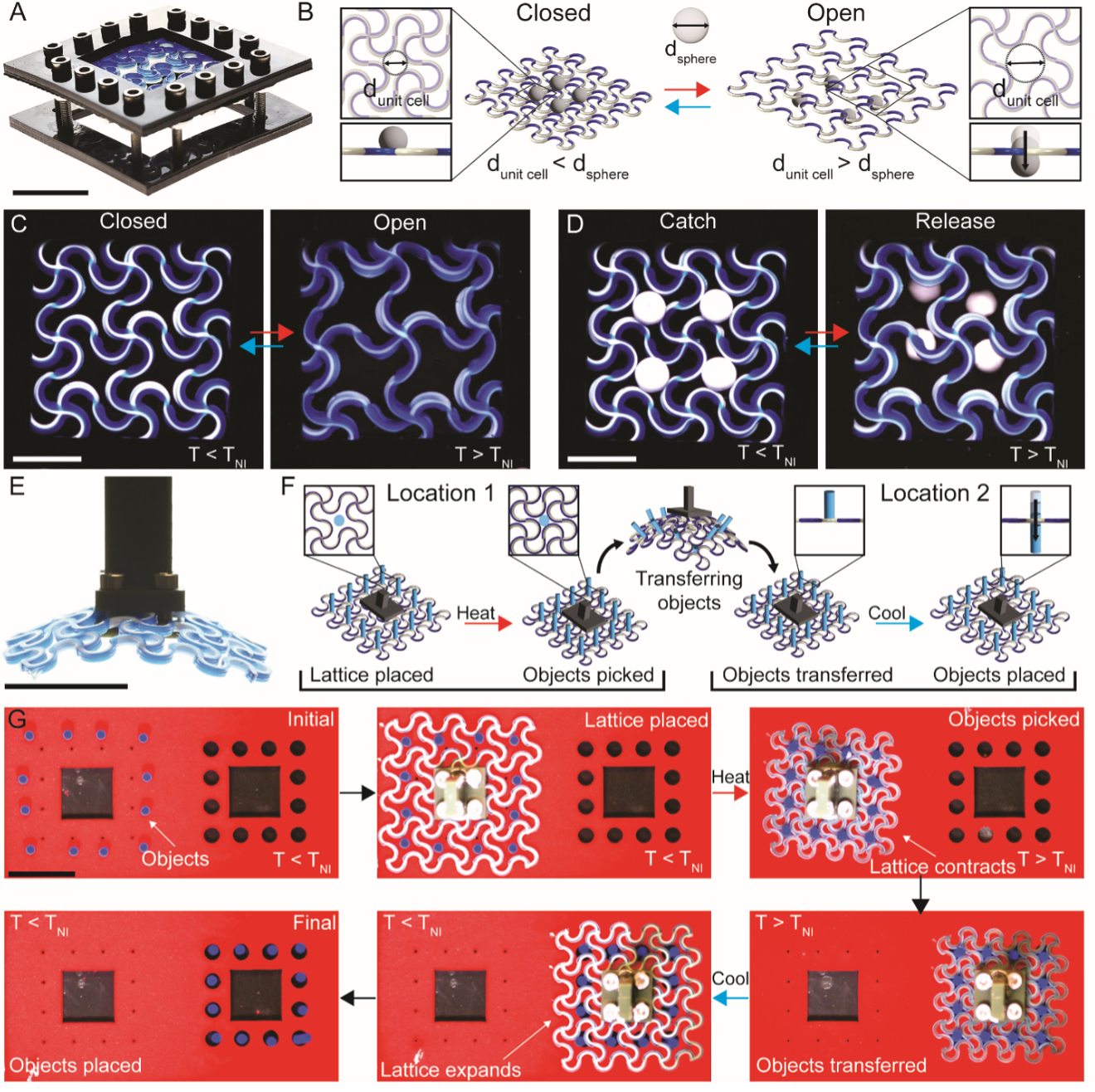}

\textbf{Figure 5. Active-passive lattices for filtering and gripping
objects. (A)} Expanding active-passive lattice filter mounted within an
acrylic frame. Scale bar: 25 mm. \textbf{(B)} Schematic of the filter
illustrating object capture at low temperature (closed state) and
release at elevated temperature (opened state). \textbf{(C)} Filter
transitioning from the closed to the opened state upon heating.
\textbf{(D)} Active-passive lattice catching and releasing spheres.
Scale bar: 10 mm for (C) and (D). \textbf{(E)} Pick-and-place gripper
composed of a contracting lattice attached to an acrylic handle. Scale
bar: 25 mm. \textbf{(F)} Schematic of the gripper illustrating object
capture upon heating and release upon cooling. \textbf{(G)} Time-lapse
images demonstrating pick-and-place multiple acrylic rods (3.5 mm
diameter, 6 mm length). Scale bar: 25 mm.

\end{document}